\documentclass[prl,twocolumn,showpacs,preprintnumbers,amsmath,amssymb]{revtex4}
\usepackage{tipa}
\usepackage{amssymb}
\usepackage{txfonts}
\usepackage{hyperref}
\usepackage{graphicx}
\usepackage{epsfig}

\begin{document}
\title{Arbitrary quantum state transfer under three parties participation$^{*}$}
\author{GUO Yan-Qing$^{\rm 1)}$$^{**}$\footnotetext{$^{*}$ Supported by Doctoral Funds of Liaoning Province of China under
Grant No. 20081077 and Funds of Dalian Maritime University under
Grant NO. DLMU-ZL-200815.\\ $^{**}$ To whom the correspondence
should be addressed as: yqguo@dlmu.edu.cn}, ZHANG Ying-Hui}
\affiliation{Department of Physics, Dalian Maritime University,
Dalian, Liaoning, 116026, P.R.China\\
} 

\begin{abstract}
Arbitrary quantum state transfer(AQST) is discussed in a system
that atoms are trapped in three separate cavities which are
connected via optical fibers. Through three parties cooperation,
the AQST can be selectively implemented deterministically. The
target state can be transferred to any of the parties with 100
percent fidelity and $\frac{1}{2}$ success probability.

\end{abstract}
\maketitle

Very recently, much attention has been paid to the study of the
possibility of quantum information processing realized via optical
fibers $^{[1,2]}$. Generating an entangled state of distant qubits
turns out to be a basic aim of quantum computation. It has been
pointed out that implementing quantum entangling gate that works
for spatially separated local processors which are connected by
quantum channels is crucial in distributed quantum computation.
Many schemes have been put forward to prepare engineering
entanglement of atoms trapped in separate optical cavities by
creating direct or indirect interaction between them $^{[3-10]}$.
Some of the schemes involve direct connection of separate cavities
via optical fibers, others apply detection of the photons leaking
from the cavities. All the implemented quantum gates work in a
probabilistic way. To improve the corresponding success
probability and fidelity, one must construct precisely controlled
coherent evolutions of the global system and weaken the affect of
photon detection inefficiency. In the system considered by
Serafini et al $^{[5]}$, the only required local control is
synchronized switching on and off of the atom-field interaction in
the distant cavities. In the scheme proposed by Mancini and Bose
$^{[11]}$, a direct interaction between two atoms trapped in
distant cavities is engineered, the only required control for
implementing quantum entangling gate is turning off the
interaction between atoms and the locally applied laser fields. In
the present letter, we propose an alternative scheme with
particular focus on the establishment of three-qubit entanglement,
which is suitable and effective for the generation of three-atom
W-type state and two-atom Bell-state. To generate three-atom
W-type state, the only control required is synchronized turning
off the locally applied laser fields. While, To generate two-atom
Bell-state, an additional quantum measurement performed on one of
the atoms is needed. We demonstrate that the scheme works in a
high success probability, and the atomic spontaneous emission does
not affect the fidelity.

The schematic setup of the system is shown in Fig. 1. Three
two-level atoms 1, 2 and 3 locate in separate optical cavities
$C_{1}$, $C_{2}$ and $C_{3}$ respectively. The cavities are
assumed to be single-sided. Three off-resonant driving external
fields $\varepsilon _{1}$, $\varepsilon _{2}$ and $\varepsilon
_{3}$ are added on $C_{1}$, $C_{2}$ and $C_{3}$ respectively. In
each cavity, a local weak laser field is applied to resonantly
interact to the atom. Two neighboring cavities are connected via
optical fiber. The global system is located in vacuum.
\begin{figure}
\epsfig{file=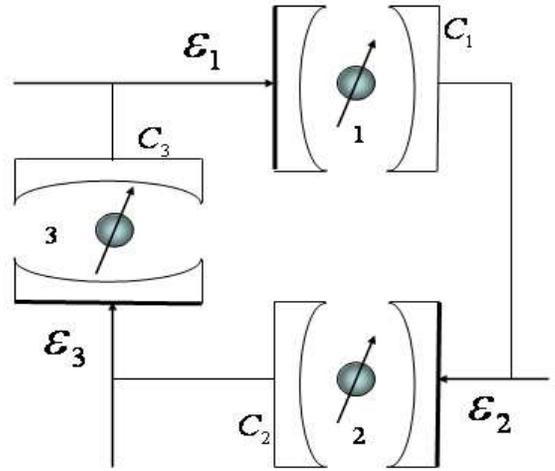, width=7.5cm,
height=6.5cm,bbllx=28,bblly=3,bburx=405,bbury=352}\caption{{\protect\footnotesize
{Schematic setup of the supposed system. Three two-level atoms are
trapped in separate optical cavities, which are connected via
optical fibers in turn. All the cavities are assumed to be
single-sided. Each of the cavities is driven by an external field.
Every atom is coupled to a local laser field.}}}
\end{figure}
Using the input-output theory, taking the adiabatic approximation
$^{[12]}$ and applying the methods developed in Refs. [11] and
[13], we obtain the effective Hamiltonian of the global system as
\begin{eqnarray}
H_{eff}=J_{12}\sigma _{1}^{z}\sigma _{2}^{z}+J_{23}\sigma
_{2}^{z}\sigma _{3}^{z}+J_{31}\sigma _{3}^{z}\sigma
_{1}^{z}+\Gamma\sum\limits_{i}(\sigma _{i}^{-}+\sigma _{i}^{+}),
\end{eqnarray}
where $\sigma_{i}^{z}$ and $\sigma_{i}^{+} (\sigma_{i}^{-})$,
$i=1,2,3$, are spin and spin raising (lowering)
operators of atom $i$, $\Gamma$ represents the local laser field
added on the atom. To keep the validity of adiabatic approximation,
we assume $\Gamma \ll J_{12}(J_{23},J_{31})$. And
\begin{eqnarray}
J_{12} &=&2\kappa\chi ^{2}Im\left\{ \alpha _{1} \alpha _{2} ^{\ast
}(Me^{i\phi
_{21}}+\kappa e^{i\phi_{32}+\phi_{13}})/(M^{3}-W^{3})\right\},  \nonumber \\
J_{23} &=&2\kappa \chi ^{2}Im\left\{ \alpha _{2} \alpha _{3}
^{\ast }(Me^{i\phi
_{32}}+\kappa e^{i\phi_{13}+\phi_{21}})/(M^{3}-W^{3})\right\},  \nonumber \\
J_{31} &=&2\kappa \chi ^{2}Im\left\{ \alpha _{3} \alpha _{1}
^{\ast }(Me^{i\phi _{13}}+\kappa
e^{i\phi_{21}+\phi_{32}})/(M^{3}-W^{3})\right\},
\end{eqnarray}
where $\kappa$ is the cavity leaking rate,
$\chi=\frac{g^2}{\Delta}$, $g$ is the coupling strength between
atom and cavity field, $\Delta$ is the detuning. In deducing Eq.
(1), the condition $\Delta\approx\kappa\gg g$ is assumed,
$M=i\Delta+\kappa$, $W^{3}=\kappa ^{3}e^{i(\phi _{21}+\phi
_{32}+\phi _{13})}$. The phase factors $\phi _{21}$, $\phi _{32}$,
and $\phi _{13}$ are the phases delay caused by the photon
transmission along the optical fibers. And
\begin{eqnarray}
\alpha_{1}&=&\frac{M^2\varepsilon_{1}+\kappa^{2}e^{i(\phi_{32}+\phi_{13})}
\varepsilon_{2}+M\kappa e^{i\phi_{13}}\varepsilon_{3}}{M^{3}-W^3},
\nonumber\\
\alpha_{2}&=&\frac{M^2\varepsilon_{2}+\kappa^{2}e^{i(\phi_{13}+\phi_{21})}
\varepsilon_{3}+M\kappa e^{i\phi_{21}}\varepsilon_{1}}{M^{3}-W^3},
\nonumber\\
\alpha_{3}&=&\frac{M^2\varepsilon_{3}+\kappa^{2}e^{i(\phi_{21}+\phi_{32})}
\varepsilon_{1}+M\kappa e^{i\phi_{32}}\varepsilon_{2}}{M^{3}-W^3},
\end{eqnarray}
 We assume that
$\varepsilon_{1}=\varepsilon_{2}=\varepsilon_{3}=\varepsilon_{0}$,
$\phi _{21}=\phi _{32}=\phi _{13}=\phi _{0}$. This leads to
\begin{eqnarray}
\alpha_{1}=\alpha_{2}=\alpha_{3}=\alpha_{0},\nonumber\\
J_{12}=J_{23}=J_{31}=J_{0}.
\end{eqnarray}
The Hamiltonian in Eq. (1) is now written as
\begin{eqnarray}
H_{eff}=H_{zz}+H_{x},
\end{eqnarray}
where
\begin{eqnarray}
H_{zz}=J_{0}(\sigma _{1}^{z}\sigma _{2}^{z}+\sigma _{2}^{z}\sigma
_{3}^{z}+\sigma _{3}^{z}\sigma _{1}^{z}),
H_{x}=\sum\limits_{i}\Gamma_{i}(\sigma _{i}^{-}+\sigma _{i}^{+}).
\end{eqnarray}
Eq. (5) represents the Hamiltonian of an Ising ring model. The
entanglement of the ground state of the above Hamiltonian has
already been discussed $^{[14]}$. Here, we study the entanglement
of the evolved system state governed by the Hamiltonian. Under the
condition $\Gamma_{i}\ll J_{0}$, the secular part of the effective
Hamiltonian can be obtained through the transformation
$UH_{x}U^{-1}$, $U=e^{-iH_{zz}t}$, as $^{[15]}$
\begin{eqnarray}
\tilde{H}=\sum\limits_{ijk}\Gamma_{i}\sigma_{i}^{x}(1-\frac{1}{2}\sigma_{j}^{z}\sigma_{k}^{z})
.
\end{eqnarray}
where the subscripts $ijk$ are permutations of $1, 2, 3$.

The straight forward interpretation of this Hamiltonian is: the
spin of an atom in the Ising ring flips \emph{if and only if} its
two neighbors have opposite spins.

For the initial states that one or two of the atoms are excited,
the system state is restricted within the subspace spanned by the
following basis vectors
\begin{eqnarray}
|\phi_{1}\rangle&=&|egg\rangle, |\phi_{2}\rangle=|eeg\rangle,
|\phi_{3}\rangle=|geg\rangle,\nonumber \\
|\phi_{4}\rangle&=&|gee\rangle, |\phi_{5}\rangle=|gge\rangle,
|\phi_{6}\rangle=|ege\rangle.
\end{eqnarray}
We firstly consider a case where $\Gamma _{1}=\Gamma_{3}=0$. The
Hamiltonian in Eq. (7) is now written as
\begin{eqnarray}
\tilde{H}=\left(\begin{array}{cccccc}
0 & \Gamma_{2} & 0 & 0 & 0 & 0\\
\Gamma_{2} & 0 & 0 & 0 & 0 & 0\\
0 & 0 & 0 & 0 & 0 & 0\\
0 & 0 & 0 & 0 & \Gamma_{2} & 0\\
0 & 0 & 0 & \Gamma_{2} & 0 & 0\\
0 & 0 & 0 & 0 & 0 & 0
\end{array}\right).
\end{eqnarray}
The eigenvalues of the Hamiltonian can be obtained as
$E_{1,2}=\pm\sqrt{2}\Gamma_{2}, E_{3,4,5,6}=0$, and the
corresponding eigenvectors are
\begin{eqnarray}
|\psi\rangle_{i}=\sum\limits_{j}S_{ij}|\phi_{j}\rangle
\end{eqnarray}
where
\begin{eqnarray}
S=\frac{1}{\sqrt{2}}\left(\begin{array}{cccccc} 1 & 1 & 0 & 0 & 0
&
0\\
-1 & 1 & 0 & 0 & 0 &
0\\
0 & 0 & 0 & 1 &
1 & 0\\
0 & 0 & 0 & -1 &
1 & 0\\
0 & 0 & 0 & 0 & 0 &
-1\\
0 & 0 & -1 & 0 & 0 & 0 \end{array}\right).
\end{eqnarray}
 For initial system state
$|\Psi(0)\rangle=\sum\limits_{i}c_{i}(0)|\phi_{i}\rangle$, the
evolving system state can be written as
$|\Psi(t)\rangle=\sum\limits_{i}c_{i}(t)|\phi_{i}\rangle$, where
the coefficients $c_{i}(t)$ are given by $^{[8]}$
\begin{eqnarray}
c_{i}(t)=\sum\limits_{j}[S^{-1}]_{ij}[Sc(0)]_{j}e^{-iE_{j}t},
\end{eqnarray}
where
$c(0)=[c_{1}(0),c_{2}(0),c_{3}(0),c_{4}(0),c_{5}(0),c_{6}(0)]^{T}$,
and $S$ is the $6\times6$ unitary transformation matrix between
eigenvectors and basis vectors.

Now we show how an arbitrary quantum state $\alpha
|e\rangle_{1}+\beta |g\rangle _{2}$ be transferred from a cavity
to another. To do this, we assume Alice, Bob and Charlie hold
atoms 1, 2 and 3 respectively, and atom 1 is supposed to be
initially in state $\alpha |e\rangle_{1}+\beta |g\rangle _{2}$,
where $\alpha$ and $\beta$ are complex numbers and fulfill
normalization condition, atoms 2 and 3 are in ground state. For
initial state $|\Psi\rangle(0)=|\phi_{1}\rangle$, one can get
\begin{eqnarray}
c_{1}(t)&=&\textrm{cos}\Gamma_{2}
t,\nonumber\\
c_{2}(t)&=&-\textrm{sin}\Gamma_{2}
t,\nonumber\\
c_{3}(t)&=&c_{4}(t)=c_{5}(t)=c_{6}(t)=0.
\end{eqnarray}

At $\Gamma_{2}t_{1}=k\pi+\frac{\pi}{2}$, Alice, Bob and Charlie
synchronously turn off driving fields $\varepsilon_{1}$,
$\varepsilon_{2}$, $\varepsilon_{3}$ and laser fields $L_{2}$, the
state of the atoms evolves to $(\alpha|ee\rangle_{12}+\beta
|gg\rangle_{12})\otimes(|g\rangle_{3})$. Now, Alice turns on her
local laser field $L_{1}$. Recall that atom 1 resonantly interacts
with $L_{1}$ and decouples from cavity field $C_{1}$. Furthermore,
since the driving fields are turned off, the Ising-type
interaction between atoms is damaged. Then, the dynamics of atom 1
is only governed effectively by the Hamiltonian
\begin{eqnarray}
H_{1}=\Gamma_{1}(\sigma ^{-}_{1}+\sigma ^{+}_{1})
\end{eqnarray}

One can get
\begin{eqnarray}
|\Psi(t)\rangle_{12}&=&|e\rangle_{1}(\alpha\textrm{cos}
\Gamma_{1}t|e\rangle_{2}-i\beta\textrm{sin}\Gamma_{1}t|g\rangle_{2})\nonumber
\\&+&|g\rangle_{1}(-i\alpha\textrm{sin}\Gamma_{1}
t|e\rangle_{2} +\beta\textrm{cos}\Gamma_{1} t|g\rangle_{2}).
\end{eqnarray}
At $\textrm{tan}\Gamma_{1} (t-t_{1})=1$, Alice performs
measurement $|e\rangle_{1}\langle e|$ on her atom, the atomic
state of Bob then is
\begin{eqnarray}|\Psi\rangle_{2}=\alpha|e\rangle_{2}-i\beta|g\rangle_{2}\end{eqnarray}. By
using a rotation

\begin{eqnarray}
H_{rot}=\left(\begin{array}{cc}
1 & 0\\
0 & e^{i\frac{\pi}{2}}
\end{array}\right).\end{eqnarray}

 Bob can obtain a state
$|\Psi\rangle_{2}=\alpha|e\rangle_{2}+\beta|g\rangle_{2}$. Thus,
Alice, Bob and Charlie cooperatively implement a perfect
deterministic quantum state transfer.

Similarly, using this method, the arbitrary quantum state can also
be transferred to Charlie.

We let $\alpha=cos(\theta)$ and $\beta=sin(\theta)$, and define
the average success probability of the quantum state transfer as
\begin{eqnarray}
P=\frac{1}{2\pi}\int^{2\pi}_{0}P(\theta)d\theta
\end{eqnarray}
the average fidelity of the quantum state transfer as
\begin{eqnarray}
F=\frac{1}{2\pi}\int^{2\pi}_{0}F(\theta)d\theta
\end{eqnarray}
where $P(\theta)=cos^{2}\theta
cos^{2}\Gamma_{1}(t-t_{1})+sin^{2}\theta
sin^{2}\Gamma_{1}(t-t_{1})$,
$F(\theta)=\frac{1}{\sqrt{P(\theta)}}(cos^{2}\theta
cos\Gamma_{1}(t-t_{1})+sin^{2}\theta sin\Gamma_{1}(t-t_{1}))$. We
can easily see that $P=\frac{1}{2}$.

\begin{figure}
\epsfig{file=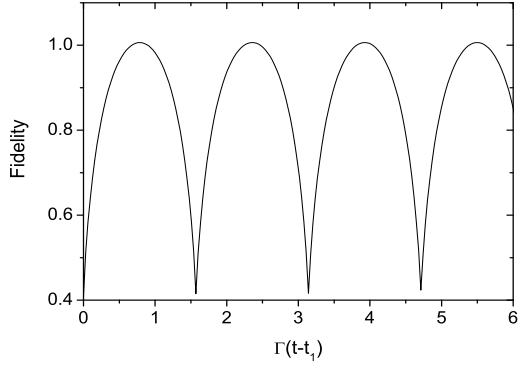, width=8cm,
height=6cm}\caption{{\protect\footnotesize {Fidelity of
transferring arbitrary quantum state from Alice to Bob with versus
time.}}}
\end{figure}
In Fig. 3, we show the fidelity $F$ with respect to time. At
$\Gamma_{1}(t-t_{1})=k\pi+\frac{3}{4}\pi$, the arbitrary quantum
state $\alpha|e\rangle+\beta|g\rangle$ can be transferred from
Alice to Bob with success probability $\frac{1}{2}$ and fidelity
100 percent.

We have put forward a scheme to transfer arbitrary quantum state
from atom to another. In this scheme, the transfer can be
deterministically implemented by three parties's cooperation. The
average success probability can approach $\frac{1}{2}$, while the
average fidelity can approach 100 percent. The transfer will be
terminated if one of the parties had a mishandling or the
communication channel had been illegally observed anywhere. So,
this scheme provides a relative more secure quantum communication
than those only using two parties. Furthermore, we can see that,
the transfer can be implemented selectively. The arbitrary quantum
state can be transferred from Alice to Bob, or from Alice to
Charlie, or from Bob to Charlie, and so on. From an extending
point of view, this kind of system may act as a perform of quantum
network.


\begin{thebibliography}{99}
\bibitem{1} Moehring D L, Maunz P, Olmschenk S, Younge K C, Matsukevich D N, Duan L M and Monroe C 2007 \textit{Nature} \textbf{449}
68

\bibitem{2} Rosenfeld W, Berner S, Volz J, Weber M and Weinfurter H 2007 \textit{Phys. Rev. Lett.} \textbf{98} 050504
\bibitem{3} Cho J and Lee H W 2005 \textit{Phys. Rev. Lett.} \textbf{95} 160501
\bibitem{4} Razavi M and Shapiro J H 2006 \textit{Phys. Rev. A} \textbf{73} 042303

\bibitem{5} Serafini A, Mancini S and Bose S 2006 \textit{Phys. Rev. Lett.} \textbf{96} 010503

\bibitem{6} Zheng S B and Guo G C 2006 \textit{Phys. Rev. A} \textbf{73} 032329

\bibitem{7} Duan L M, Madsen M J, Moehring D L, Maunz P, Kohn R N and Monroe C 2006 \textit{Phys. Rev. A} \textbf{73} 062324

\bibitem{8} Yin Z Q and Li F L 2007 \textit{Phys. Rev. A} \textbf{75} 012324

\bibitem{9} Lu D M and Zheng S B 2007 \textit{Chin. Phys. Lett.} \textbf{24} 596

\bibitem{10} Ou Y C, Yuan C H and Zhang Z M 2006 \textit{J. Phys. B: At. Mol. Opt. Phys.} \textbf{39} 7

\bibitem{11} Mancini S and Bose S 2004 \textit{Phys. Rev. A} \textbf{70}
022307

\bibitem{12} Walls D F and Milburn G J 1994 \textit{Quantum Optics} (Berlin: Springer)chap 7 p121

\bibitem{13} Guo Y Q, Chen J and Song H S 2006 \textit{Chin. Phys.
Lett.} \textbf{23} 1088

\bibitem{14} \v{S}temlmachovi\v{c} P and Bu\v{z}ek V 2004 \textit{Phys. Rev. A} \textbf{70}
032313

\bibitem{15} Lee J S and Khitrin A K 2005 \textit{Phys. Rev. A} \textbf{71} 062338

\bibitem{16} Coffman V, Kundu J and Wootters W K 2000 \textit{Phys. Rev.
A} \textbf{61} 052306

\bibitem{17} Wootters W K 1998 \textit{Phys. Rev. Lett.} \textbf{80} 2245

\bibitem{18} Tittel W, Brendel J, Gisin B, Herzog T, Zbinden H and Gisin N 1998 \textit{Phys. Rve. A} \textbf{57} 3229


\end{thebibliography}
\end{document}